# PRO: Projection Domain Synthesis for CT Imaging

Kang Chen, Bin Huang, Xuebin Yang, Junyan Zhang, Yongbo Wang, Qiegen Liu, *Senior Member, IEEE*

***Abstract*—Synthetic CT projection data is crucial for advancing imaging research, yet its generation remains challenging. Current image domain methods are limited as they cannot simulate the physical acquisition process or utilize the complete statistical information present in projection data, restricting their utility and fidelity. In this work, we present PRO, a projection domain synthesis foundation model for CT imaging. To the best of our knowledge, this is the first study that performs CT synthesis in the projection domain. Unlike previous approaches that operate in the image domain, PRO learns rich structural representations from projection data and leverages anatomical text prompts for controllable synthesis. Projection data generation models can utilize complete measurement signals and simulate the physical processes of scanning, including material attenuation characteristics, beam hardening, scattering, and projection geometry, and support research on downstream imaging tasks. Moreover, PRO functions as a foundation model, capable of generalizing across diverse downstream tasks by adjusting its generative behavior via prompt inputs. Experimental results demonstrated that incorporating our synthesized data significantly improves performance across multiple downstream tasks, including low-dose and sparse-view reconstruction. These findings underscore the versatility and scalability of PRO in data generation for various CT applications. These results highlight the potential of projection domain synthesis as a powerful tool for data augmentation and robust CT imaging. Our source code is publicly available at: https://github.com/yqx7150/PRO.**

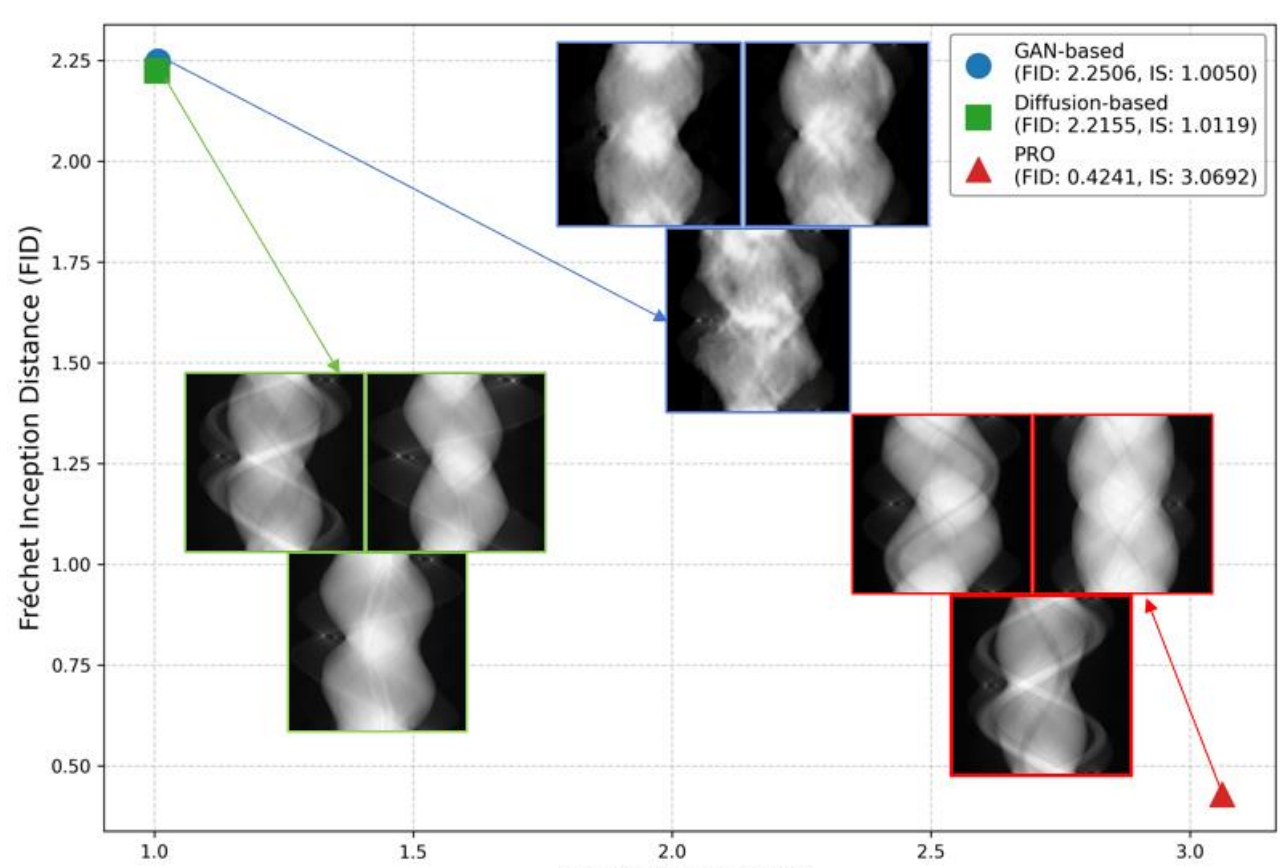

**Fig. 1.** Comparison of FID and IS metrics for 1000 CT data generating from projection domain.

## I. Introduction

Deep learning has driven significant advances in medical image analysis, including tasks such as image segmentation [1], reconstruction [2], inpainting [3], and modality translation [4]. However, its success critically depends on the availability of large scale, high quality annotated datasets. Unlike general computer vision, medical image datasets are often small due to data acquisition costs, annotation difficulty, privacy concerns, and the low prevalence of certain diseases. This data scarcity severely limits the performance and generalizability of deep learning models in real world clinical applications.

Generating synthetic data with privacy guarantees provides a promising alternative, allowing meaningful research to be carried out at scale [5-7]. Together with traditional data augmentation techniques e.g., geometric transformation, these synthetic data could complement real data to dramatically increase the training set of machine learning models. generative models learn the probability density function underlying the data they are trained on and can create realistic representations of examples which are different from the ones present in the training data by sampling from the learned distribution. However, generating meaningful synthetic data is not easy, especially when considering complex medical information.

Nowadays, generative adversarial networks (GANs) have been applied in various fields to generate synthetic images, producing realistic and clear images and achieving impressive performance [8, 9]. For example, [10] combined variational autoencoders with GANs to generate various modalities of whole brain volumes from a small training set and achieved a better performance compared to several baselines. However, since their study resized the images to a small volume before training, with a size of 64 × 64 × 64 voxels, their synthetic medical images did not replicate many essential finer details. In addition, due to the prevalence of 3D high resolution data in the field, researchers tended to have their models restrained by the amount of GPU memory available. To mitigate this issue, [11] proposed a 3D GAN with a hierarchical structure which could generate a low resolution version of the image and anchor the generation of high resolution sub volumes on it. With this approach, the authors were able to generate impressive realistic 3D thorax CT and brain MRI with resolutions up to 256 × 256 × 256 voxels. Despite attracting great interest, GANs still come with inherent challenges, such as being notoriously unstable during training and failing to converge or to capture the variability of the generated data due to mode collapse issues [12].

This work was supported National Key Research and Development Program of China (2022YFA1204200) and in part by the Guangdong Basic and Applied Basic Research Foundation (2023A1515011780). (Kang Chen and Bin Huang are co-first authors.) (Corresponding authors: Yongbo Wang and Qiegen Liu.)

This work did not involve human subjects or animals in its research.

K. Chen, B. Huang, X. Yang and J. Zhang are with School of Mathematics and Computer Sciences, Nanchang University, Nanchang 330031, China ({356100240002, huangbin, ice, zhangjunyan}@email.ncu.edu.cn).

Q. Liu is with School of Information Engineering, Nanchang University, Nanchang 330031, China (liuqiegen@ncu.edu.cn).

Y. Wang is with the Key Laboratory of Biomedical Information Engineering of Ministry of Education, School of Life Science and Technology, Xi'an Jiaotong University, Xi'an 710049, China (wyb081@xjtu.edu.cn).

Diffusion model has recently emerged as a compelling alternative due to their strong theoretical foundation and superior performance in sample quality and diversity. By iteratively denoising samples from a Gaussian prior, diffusion model offers a more stable and controllable generation process. Latent diffusion model (LDM) [13] further extended this framework by compressing images into latent spaces for efficient training and inference, enabling high resolution synthesis at significantly reduced computational cost. For example, Pinaya *et al*. proposed a 3D latent diffusion model that could generate realistic brain MRI volumes conditioned on semantic labels, achieving competitive performance in structure preserving synthesis [14]. Guo *et al*. introduced MAISI, a multi-task medical image synthesis framework that integrated LDMs and ControlNet to support high fidelity CT generation guided by clinical prompts and organ masks [15]. Additionally, Generatect was capable of synthesizing entire CT volumes from textual input [16], pushing the boundary of multi-modal conditional generation in clinical contexts.

Despite these advances, existing generative models almost exclusively operate in the image domain, where final reconstructed images or latent features are directly synthesized. While this approach aligns with the format of downstream tasks, it neglects the easurement process that lies at the core of tomographic imaging. CT projection data acquired in Radon space, encodes rich physical priors, such as cross detector correlations, material attenuation patterns, and view dependent anatomical structure encoding. These physical characteristics are typically degraded or lost during the reconstruction process. As a result, ignoring the projection domain may lead to generated images that lack realism and fail to accurately reflect the physics of the imaging system.

To bridge this gap, we propose PRO, a novel diffusion based generative framework that performs CT image synthesis directly in the projection domain. Unlike conventional image domain approaches, PRO learns to generate raw projection data, thereby modeling acquisition level signal properties such as cross detector correlations, material attenuation patterns, and view dependent anatomical structure encoding. These characteristics are naturally embedded in the projection space while typically diminished or distorted in reconstructed images. In fact, our research group have developed several diffusion models in projection domain for various CT reconstruction tasks, GMSD [17], OSDM [18], RAP [19], SWORD [20], etc. Generative priors learned directly from projection domain exhibits excellent characteristics and subsequent reconstruction performance.

In this work, we take a significant step further by introducing a prompt driven foundation model for CT Synthesis in projection domain, capable of adapting to diverse downstream tasks by conditioning its generation behavior on task specific textual prompts. These prompts guide the model to generate projection data that reflects corresponding acquisition conditions. This flexible control allows PRO to serve as a unified foundation model, enabling data synthesis tailored to various CT imaging scenarios. These synthesized projections are then reconstructed into CT images using filtered back projection (FBP) algorithm, resulting in anatomically consistent and diverse outputs. While projection domain synthesis offers clear advantages, Fig. 1 highlights the inherent difficulty of producing high quality projection data. Traditional generative models such as diffusion-based and GAN-based frameworks struggle to generate structurally coherent sinograms, whereas PRO demonstrates superior fidelity. Minor artifacts or noise present in the sinogram can be dramatically magnified during the reconstruction process, resulting in widespread degradation across the image. To enhance the image quality, especially local structures and noise characteristics, we apply a lightweight CNN-based image domain refinement module, SharpNet, after the initial reconstruction. Extensive experiments across multiple tasks, including low-dose CT reconstruction and sparse-view reconstruction, demonstrate that incorporating projection domain synthesized data generated by PRO leads to substantial performance improvements. These results validate the utility of projection domain synthesis as an effective strategy for data augmentation and training enhancement, particularly when labeled data are limited. The main contributions of this work are summarized as follows:

- To the best of our knowledge, PRO is the first work to explore CT image synthesis in the projection domain using diffusion model, enabling more faithful modeling of acquisition physics and anatomical structures. PRO is guided by task specific textual prompts for flexible adaptation to various CT imaging scenarios.
- Incorporating synthetic data of PRO significantly boosts performance in multiple downstream tasks such as low-dose and sparse-view reconstruction. These findings illustrate that projection domain generation can act as a task adaptive foundation framework under varied clinical conditions.

The rest of the study is organized as follows: Relevant background on diffusion model is described in Section II. Detailed procedure and algorithm are presented in Section III. Experimental results and specifications about the implementation are given in Section IV. We conclude our work in Section V and Section VI, respectively.

## II. Preliminary

### *A. Deep Generative Models*

Recently advances with deep generative networks have shown obvious gains in modeling complex distributions such as images [21], audios [22] and texts [23]. The popular deep generative models can be primarily categorized into two groups, explicit generative model and implicit generative model. The former model provides an explicit parametric specification of the data distribution, including autoencoders (AE) and its variants [24], [25], flow-based generative models [26], [27], score-based model [28] and deep Boltzmann machine [29]. Specially, for a log likelihood function $\log p(x)$, score-based model train parametric network to approximate the likelihood gradient $\nabla_x \log p(x)$. Alternatively, we can specify implicit probabilistic model that defines a stochastic procedure to directly generate data. GANs [30] are the well-known implicit likelihood models. They optimize the objective function using adversarial learning and have been shown to produce high quality images [31]. Due to the occurrence of pattern collapse, GANs still suffer from a remarkable difficulty in training.

### *B. Denoising Diffusion Probabilistic Model*

Diffusion model has emerged as frontrunners in both density estimation [23] and sample quality enhancement [24]. These models utilize parameterized Markov chains to optimize the lower variational bound on the likelihood function, enabling them to generate target distributions with greater accuracy compared to alternative generative models.

The diffusion process operates on an input image $x_0$, gradually transforming it into Gaussian noise $x_t \sim \mathcal{N}(0,\mathbf{I})$ through $t$ iterations as shown in Fig. 2. Each iteration of this process is described as follows:

$$q(x_t | x_{t-1}) = \mathcal{N}(x_t; \sqrt{1-\beta_t} x_{t-1}, \beta_t \mathbf{I}) \tag{1}$$

where $x_t$ denotes the noised image at time step $t$, $\beta_t$ represents the predefined scale factor, and $\mathcal{N}$ represents the Gaussian distribution. During the reverse process, diffusion model samples a Gaussian random noise map $x_t$, then progressively denoise $x_t$ until it achieves a high quality output $x_0$:

$$p(x_{t-1} | x_t, x_0) = \mathcal{N}(x_{t-1}; \mu_t(x_t, x_0), \sigma_t^2 \mathbf{I}) \tag{2}$$

To train a denoising network $\epsilon_\theta(x_t, t)$, given a clean image $x_0$, diffusion model randomly samples a time step $t$ and a noise $\epsilon \sim \mathcal{N}(0,\mathbf{I})$ to generate noisy images $x_t$ according to Eq. (2).

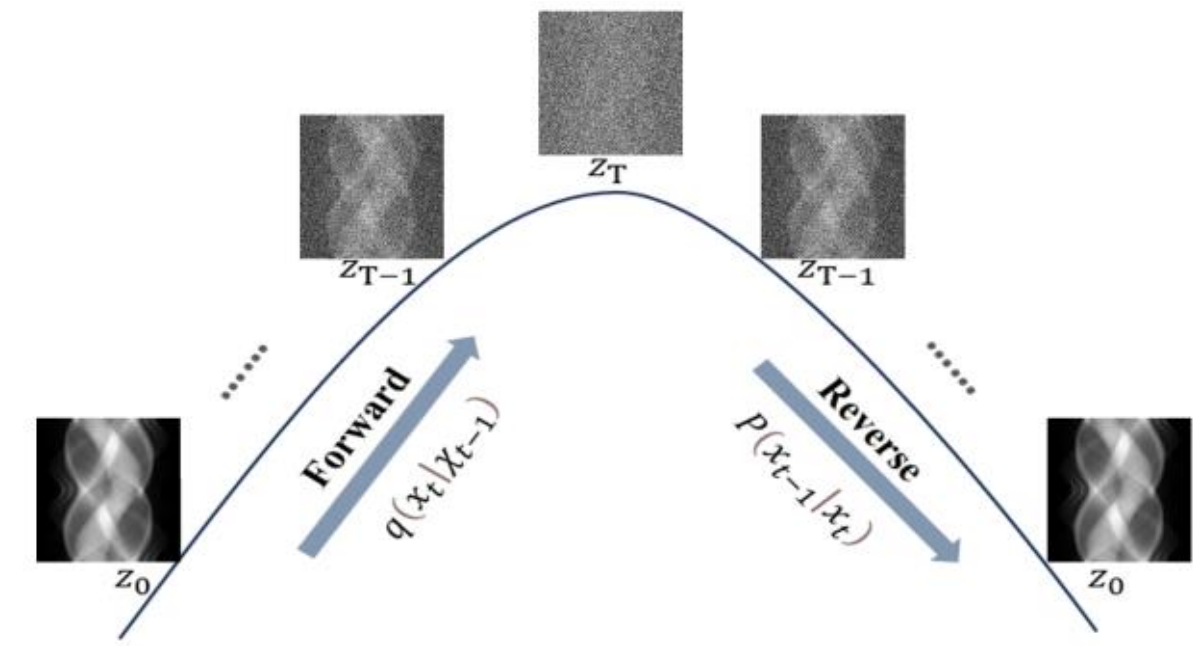

**Fig. 2.** Forward and reverse processes of DDPM.

### C. Score-based Diffusion Model

Score-based diffusion model perturb the data distribution according to the forward SDE by injecting Gaussian noise, arriving at a tractable distribution, such as an isotropic Gaussian distribution. To sample data from this distribution, one can train a neural network to estimate the gradient of the log data distribution, namely the score function $\nabla_x \log p(x)$, which can be used to solve the reverse SDE numerically.

The diffusion process can be modeled as the solution to the following SDE:

$$dx = f(x,t)dt + g(t)dw \tag{3}$$

where $f(x,t) \in \mathbb{R}^n$ and $g(t) \in \mathbb{R}$ correspond to the drift and diffusion coefficients, respectively. $w \in \mathbb{R}^n$ induces the Brownian motion.

One can construct different SDEs by choosing different functions $f(x,t)$ and $g(t)$. Since variance exploding (VE) SDEs can lead to higher sample qualities, here we develop the following VE-SDE:

$$f(x,t) = 0,\ g(t) = \sqrt{d[\sigma^2(t)]/dt} \tag{4}$$

where $\sigma(t) > 0$ is a monotonically increasing function, which is typically chosen to be a geometric series [32], [33]. It is well known that the reverse of a diffusion process is also a diffusion process. Thus, we obtain the following reverse-time SDE:

$$\begin{aligned} dx &= [f(x,t) - g(t)^2 \nabla_x \log p_t(x)]dt + g(t)d\bar{w} \\ &= (d[\sigma^2(t)]/dt)\nabla_x \log p_t(x) + \sqrt{(d[\sigma^2(t)]/dt)}d\bar{w} \end{aligned} \tag{5}$$

where $\bar{w}$ is a standard Brownian motion with time flows backwards from $T$ to $0$, and $dt$ is an infinitesimal negative timestep. Once the score of each marginal distribution $\nabla_x \log p_t(x)$ is known for all $t$, we can derive the reverse diffusion process from Eq. (3) and simulate it to sample from $p_0$.

In order to solve Eq. (5), one has to know the score function for all $t$, which can be estimated using a time conditional neural network $S_\theta(x,t) \simeq \nabla_x \log p_t(x(t))$. The training parameters $\theta$ includes all the weights and bias raised in interpolation and denoising networks. Since we do not know the true score, we can use denoising score matching [32] to replace the unknown $\nabla_x \log p_t(x)$ with $\nabla_x \log p_t(x(t) | x(0))$, where $\nabla_x \log p_t(x(t) | x(0))$ is the Gaussian perturbation kernel centered at $x(0)$. Under some regularity conditions, $S_\theta(x,t)$ trained with denoising score matching will satisfy $S_{\theta^*}(x,t) = \nabla_x \log p_t(x(t))$ almost surely [28]. Formally, we optimize the parameters $\theta$ of the score network with the following cost:

$$\begin{aligned} \theta^* = \arg\min_\theta \mathbb{E}_t \{ \lambda(t) \mathbb{E}_{x(0)} \mathbb{E}_{x(t)|x(0)} \\ [\| S_\theta(x(t),t) - \nabla_{x(t)} \log p_t(x(t) | x(0)) \|^2 ] \} \end{aligned} \tag{6}$$

Once the network is trained with Eq. (6), we can plug the approximation $S_\theta(x,t) \simeq \nabla_x \log p_t(x(t))$ to solve the reverse SDE in Eq. (5):

$$dx = (d[\sigma^2(t)]/dt) S_\theta(x,t) + \sqrt{(d[\sigma^2(t)]/dt)}d\bar{w} \tag{7}$$

Then, we can solve the SDE numerically with Euler discretization [34]. This involves discretizing $t$ in range [0,1], which is uniformly separated into $N$ intervals such that $0 = t_0 < \cdots < t_N = 1$, $\Delta t = 1/N$. Additionally, we can correct the direction of gradient ascent with Langevin Markov Chain Monte Carlo algorithm [33]. Iteratively implementing predictor and corrector step yields the predictor corrector sampling algorithm [34].

## III. Proposed Method

### A. Motivation

Existing methods that generate data in the image domain suffer from fundamental limitations. They cannot accurately simulate the physical data acquisition process, which restricts their applicability in CT imaging research. Moreover, such approaches are unable to fully exploit the complete measurement information available in projection data, thereby limiting the realism and effectiveness of the synthesized output. In contrast, generating data directly in the projection domain offers a more powerful alternative. This strategy leverages the complete raw measurement signals to achieve higher fidelity synthesis and explicitly incorporates key physical aspects of the imaging process, including material-dependent attenuation, beam hardening, scatter effects, and system geometry. As a result, it provides realistic sinogram data that can effectively support a wide range of downstream imaging tasks such as reconstruction, artifact correction, and protocol optimization. Although [35] has applied LDM as the backbone in the projection domain, it only fits the inpainting problems.

[36] focus on solving different completion problems as an inpainting based foundation model in the projection domain. Both of them rely on conditional inpainting within partially observed sinograms, rather than modeling the full distribution of projection data without prior measurements.

This study presents PRO, the first framework to synthesize realistic CT projection data entirely without prior measurements. By enabling prompt based control, PRO supports diverse CT imaging scenarios and functions as a unified foundation model capable of adapting its generative behavior across different tasks. This approach not only increases generalizability but also unlocks controllable data synthesis in the projection domain. Despite its advantages, generating high quality projection data is inherently challenging. Even subtle noise or artifacts in the sinogram can be significantly amplified during image reconstruction, leading to global distortions. To address this, we build a diffusion model in projection domain (DMPD) upon a LDM backbone and introduce a multi-latent space architecture, where different latent spaces correspond to different prompts. This design enhances the model's ability to learn structured projection features under varied task conditions and improves generation accuracy across tasks. To further refine the synthesized outputs, a lightweight image domain enhancement module is proposed called SharpNet. SharpNet is designed to suppress residual noise and enhance fine grained structures by post processing in image domain, effectively improving anatomical clarity and visual fidelity in the final images. In the following sections, we describe DMPD, the text conditioning mechanism, and SharpNet in detail.

*B. DMPD*

To overcome the limitations of conventional image domain generation approaches and exploit the structurally, physically and meaningful information embedded in projection data, we propose PRO, shown in Fig. 3, a novel latent diffusion framework that operates directly in the projection domain. This approach allows the model to learn low level acquisition patterns such as raw attenuation profiles and cross detector correlations, which are critical for capturing anatomical fidelity but often degraded during image reconstruction.

Instead of synthesizing images in pixel level space, we first encode projection data into a perceptual latent space using a projection domain encoder $\varepsilon_{proj}$ and reconstruct it through a decoder $D_{proj}(z)$. This perceptual compression stage allows the diffusion model to operate efficiently on a compact yet informative representation:

$$z = \varepsilon_{proj}(x_{proj}) \text{ , } \tilde{x}_{proj} = D_{proj}(z) \tag{8}$$

where $x_{proj} \in \mathbb{R}^{H \times W}$ is the raw projection data, and $z \in \mathbb{R}^{h \times w \times c}$ is the latent projection embedding. The encoder downsamples the input by a factor $f$, and $h = H / f$, $w = W / f$.

To regularize the learned latent space and maintain anatomical consistency, we explore two variants: (i) a KL regularized latent diffusion variant $\mathcal{L}_{KL}$, and (ii) a quantized version $\mathcal{L}_{VQ}$ using a vector quantization layer embedded in the decoder. The KL regularized form aligns the latent distribution with a standard normal prior, while the quantized version maintains structural coherence by constraining the latent codes to a learned dictionary.

Following compression, we perform text guided generation in this latent projection space. We formulate the denoising objective over latent projection representations as:

$$\mathcal{L}_{proj} = \mathbb{E}_{z_t,y}[\| \varepsilon_\theta(z_t,t,\tau(y)) - \epsilon \|^2] \tag{9}$$

where $z_t \sim \mathcal{N}(0,I)$ is a noisy latent, $\tau(y)$ is the textual embedding from anatomical prompts, and $\epsilon$ denotes the ground truth noise used for training. The denoising network $\varepsilon_\theta$ is implemented as a U-Net backbone conditioned on cross modal text features using cross attention layers.

The forward diffusion $q$ that gradually corrupts $z_0$ into $z_t$ over $T$ time steps:

$$q(z_t \mid z_0) = \mathcal{N}(z_t; \sqrt{\bar{\alpha}_t} z_0, (1-\bar{\alpha}_t)\mathbf{I}) \tag{10}$$

where $\bar{\alpha}_t = \prod_{i=1}^{t} \alpha_i$ and $\alpha_i \in (0,1)$ is a noise schedule. During training, the model learns to predict the noise $\epsilon$ added at each step via a conditional U-Net denoiser $\epsilon_\theta$, using both the noised latent $z_t$ and the text embedding $\tau(y)$ as inputs:

$$\mathcal{L}_{diff} = \mathbb{E}_{z_0,t,\epsilon}\left[\| \epsilon - \epsilon_\theta(z_t,t,\tau(y)) \|^2\right] \tag{11}$$

At inference stage, we generate samples by reverse diffusion, starting from $\mathbf{z}_T \sim \mathcal{N}(0,\mathbf{I})$, and iteratively applying:

$$z_{t-1} = (1/\sqrt{\alpha_t})(z_t - ((1-\alpha_t)/(\sqrt{1-\bar{\alpha}_t}))\epsilon_\theta(z_t,t,\tau(y))) + \sigma_t \cdot \eta_t \tag{12}$$

where $\eta_t \sim \mathcal{N}(0,\mathbf{I})$, and $\sigma_t$ controls the stochasticity.

The denoised latent $z_0$ is decoded back to the projection domain using a decoder $D_p$:

$$\hat{x}_{\text{proj}} = D_p(z_0) \tag{13}$$

This enables projection to image conversion via traditional methods or learning based refinements.

Upon generation of the denoised latent $\hat{z}_0$, we generate the projection data via $D_{proj}(\hat{z}_0)$, and subsequently apply FBP to obtain the corresponding synthetic medical image.

Our framework thus separates generative modeling in the projection domain from visual refinement in the image domain, leading to higher anatomical accuracy, reduced training data requirements, and enhanced generalization across body regions.

*C. Text Conditioning with Task Specific Latent Spaces*

To enable text driven generation tailored to diverse CT imaging tasks, we adopt a task specific latent space conditioning mechanism, where each input text prompt $y_i$ dynamically determines the corresponding latent space $\mathcal{Z}_i$ in which diffusion occurs.

Let $y_i$ denote a task specific text prompt, such as "brain" or "body", these prompts are first embedded into a semantic representation via a text encoder:

$$\tau(y_i) = TextEncoder(y_i) \tag{14}$$

Unlike conventional LDMs where all prompts operate in a shared latent space, we define a set of task specific latent spaces $\{\mathcal{Z}_i\}_{i=1}^{N}$, each tailored to a specific class of medical generation tasks. The prompt embedding $\tau(y_i)$ is used to select or activate the corresponding latent space $\mathcal{Z}_i$, enabling precise modeling of the underlying projection distribution.

Within each $\mathcal{Z}_i$, the denoising process is conditioned on both time step t and the task relevant text embedding:

$$\hat{\epsilon}_{i,t} = \varepsilon_{\theta i}(z_{i,t}, t, \tau(y_i)) \text{ , for } z_{i,t} \in \mathcal{Z}_i \tag{15}$$

where $\varepsilon_{\theta i}$ is the noise prediction network associated with task $i$, and $z_{i,t}$ is a noisy latent vector sampled from the task specific latent trajectory in $\mathcal{Z}_i$. This formulation ensures that the denoising process is semantically and structurally aligned with the type of CT scan indicated by $y_i$.

Optionally, if memory sharing is desired, the denoising model $\varepsilon_{\theta i}$ may share a global backbone and employ conditional adapters or cross attention mechanisms modulated by $\tau(y_i)$, allowing soft parameter sharing while retaining task specific expressivity.

This task guided latent space conditioning mechanism significantly enhances the model's ability to generate projection domain data that accurately reflects anatomical priors and distribution patterns described in the text.

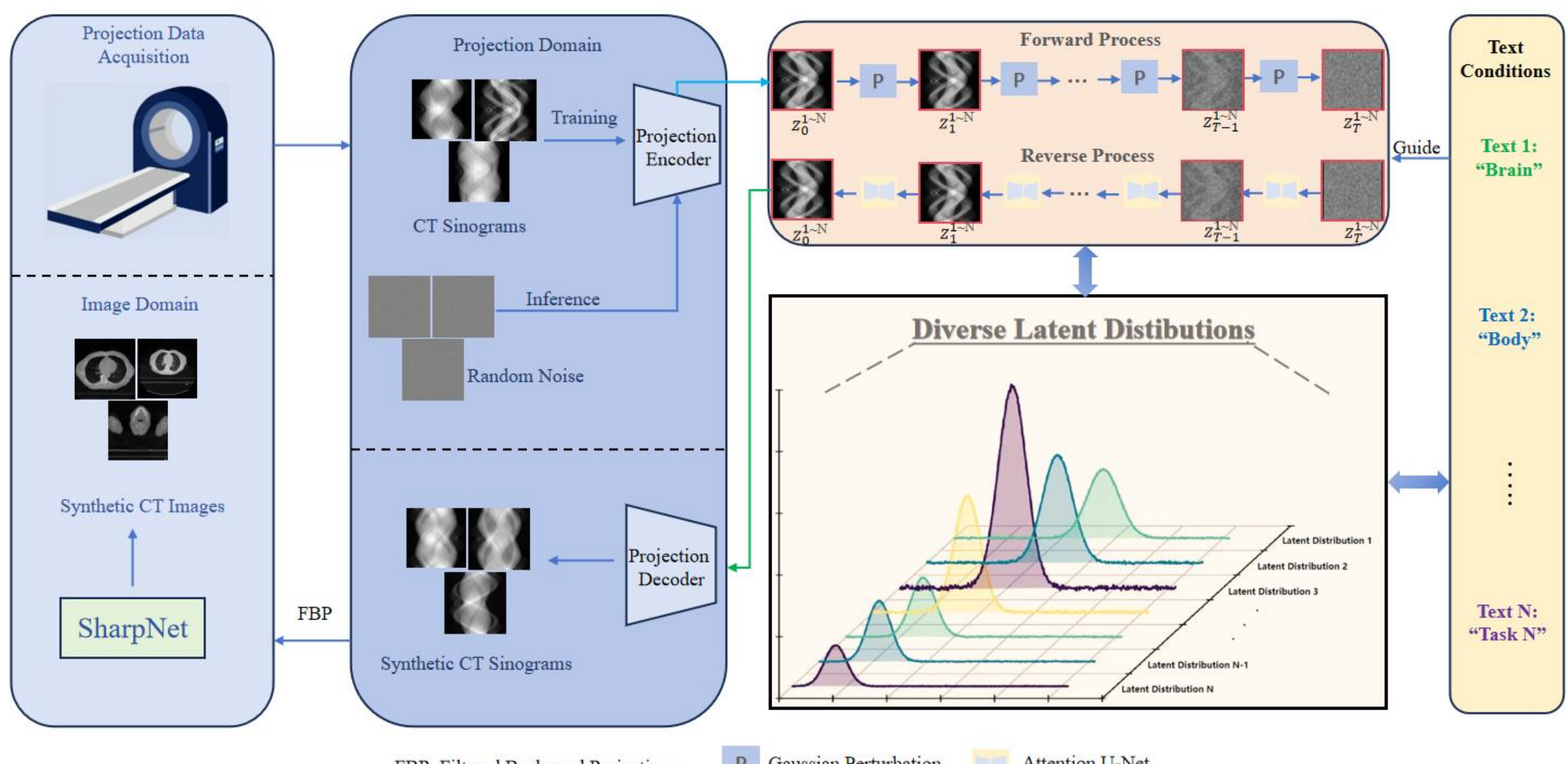

**Fig. 3.** The pipeline of the PRO framework. PRO mainly consists of DMPD and SharpNet. The text prompt is first encoded and used to select a task specific latent space. In this latent space, PRO performs diffusion generation to produce projection data. The generated projection is then converted to CT images using FBP algorithm. SharpNet refines CT images by enhancing details and removing noise, producing high quality medical images aligned with the given text prompt.

### *D. SharpNet*

While projection domain generation preserves physical consistency and structural patterns, it often lacks high frequency details due to the inherently lossy nature of both projection encoding and diffusion sampling. To mitigate this, we introduce SharpNet as shown in Fig. 4, a dedicated image domain refinement module that enhances the anatomical realism and sharpness of the final reconstructed images.

SharpNet is a lightweight CNN-based post processing network trained to map coarse reconstructed images $x_{coarse}$ (obtained by the generated projection data using FBP to sharper, high fidelity outputs $x_{refined}$:

$$I_{refined} = SharpNet(I_{coarse}),\ I_{coarse} = FBP(\hat{x}_{proj}) \tag{16}$$

Architecturally, SharpNet adopts a U-Net like structure with multi-scale skip connections, specifically designed to correct artifacts and hallucinations introduced during the projection to image inversion process. Its encoder captures context aware features at multiple scales, while the decoder restores local structural details through residual refinement pathways. Unlike DMPD, which is a generative model trained in a latent diffusion framework, SharpNet is a supervised CNN based network. However, due to the lack of fully paired image datasets, we simulate realistic supervisory signals by injecting Gaussian noise ($\sigma$ = 0.1) into the image space, thereby synthesizing pseudo ground truth image $I_{gt}$ and noisy image $I_{noisy}$ pairs. This strategy enables SharpNet to learn denoising and detail enhancement mappings effectively, even in the absence of strictly aligned training pairs.

We use the Mean Squared Error (MSE) loss and the multi-scale perceptual loss to reduce the excessive smoothness of the output image and decrease the image blurriness.

$$\mathcal{L}_{\text{mse}} = \left\| F(I_{noisy}, \theta) - I_{gt} \right\|^2 \tag{17}$$

$$\mathcal{L}_{\text{multi-p}} = (1/S)\sum_{s=1}^{S} \left\| \varphi_s(F(I_{noisy}, \theta), \hat{\theta}) - \varphi_s(I_{gt}, \hat{\theta}) \right\|^2 \tag{18}$$

where $I_{noisy}$ serves as the input, $I_{gt}$ serves as the target, $N$ is the number of images. $F$ represents a denoising model with parameters $\theta$. The symbol $\varphi$ represents a model with fixed pre-trained weights $\hat{\theta}$, which is used to calculate the perceptual loss. $S$ is the number of scales.

In summary, the total loss of SharpNet is:

$$\mathcal{L}_{Sharp} = \mathcal{L}_{\text{mse}} + \lambda_1 \mathcal{L}_{\text{multi-p}} \tag{19}$$

Optionally, SharpNet can be extended to incorporate uncertainty aware enhancement using dropout layers at inference, or can be conditioned on the same text prompt $\tau(y)$ to encourage semantically guided refinement in downstream tasks such as lesion highlighting or tracer specific denoising.

Through the collaborative operation of DMPD and SharpNet, our framework achieves both physically grounded synthesis in the projection domain and visually compelling rendering in the image domain. The total objective function of PRO is:

$$\mathcal{L}_{total} = \mathcal{L}_{proj} + \lambda_2 \mathcal{L}_{Sharp} \tag{20}$$

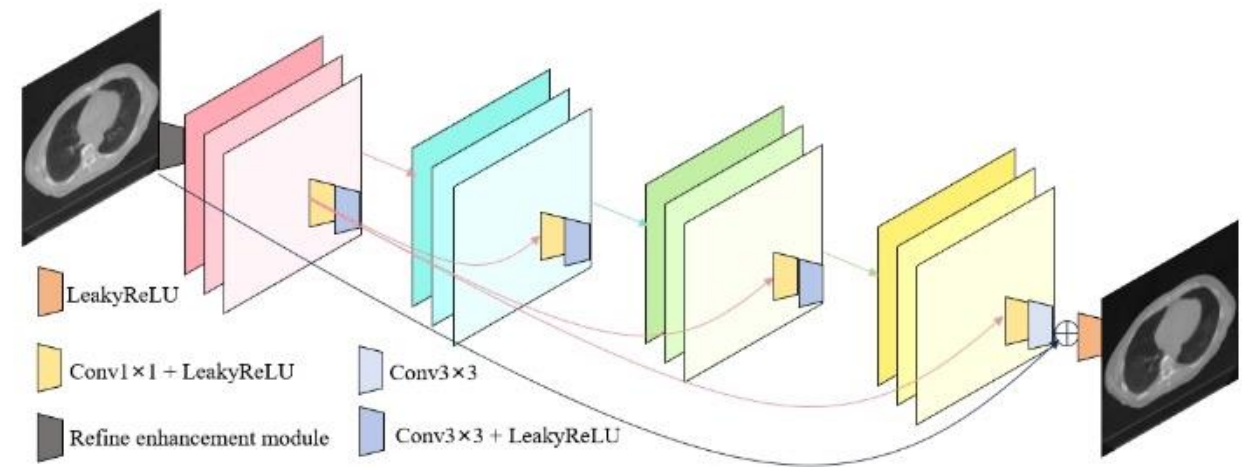

**Fig. 4.** Pipeline of the SharpNet refinement module. SharpNet is a lightweight CNN that refines coarse medical images reconstructed from projections. It uses a U-Net like structure to remove noise and recover fine details, guided by simulated noisy clean image pairs and trained with MSE and perceptual loss.

Algorithm 1 represents the overall training and inference process.

**Algorithm 1: Training and inference stages**

**Training stage**

**Dataset:** $x_{gt} \in \mathbb{R}^{H\times W}$ **,** $\tau(y_i)$ **.**
**1: Repeat**
**2: Sample** $x_{gt}, t \sim D$
**3: Encode projection to latent:** $z_0 = \varepsilon_{proj}(x_{proj})$
**4: Text embedding:** $\tau(y_i) = TextEncoder(y_i)$
**5: Predict noise:** $\hat{\epsilon}_{i,t} = \varepsilon_{\theta i}(z_{i,t}, t, \tau(y_i))$
**6: Generate Sinogram:** $\hat{x} = D(z_0)$
**7: Loss:** $\mathcal{L}_{total} = \mathcal{L}_{proj} + \lambda_2 \mathcal{L}_{Sharp}$
**8: Take gradient descent:** $\nabla_\theta \mathcal{L}_{total}$
**9: Until convergence**
**10:Return trained model** $\epsilon_\theta$ **, encoder** $\varepsilon_{proj}$ **, decoder** $D$ **, and refiner** *SharpNet*

**Inference stage**

**Setting: Text prompt** $y_i$
**1: Initialize latent** $z_T \sim \mathcal{N}(0, I)$
**2: Text embedding** $\tau(y_i) = TextEncoder(y_i)$
**3: For** $t = T$ **to 1**
**4: Predict noise:** $\hat{\epsilon} = \epsilon_\theta(z_t, t, c_t)$
**5: Update latent:** $z_{t-1}$ **updated by Eq. 9**
**6: End for**
**7: Decode image:** $\hat{x} = D(z_0)$ , $I_{coarse} = FBP(\hat{x}_{proj})$
**8: Refine image:** $I_{refined} = SharpNet(I_{coarse})$
**9: Return final image:** $I_{refined}$

## IV. EXPERIMENTS

### A. Data Specification

The evaluation employs the simulated data of human abdominal images provided by the Mayo Clinic for the AAPM Low Dose CT Grand Challenge [37]. The data includes high-dose CT scans of 10 patients. For the training of the DMPD, 12,450 slices are used. Each slice covers an area of 512×512 pixels. Among them, 11274 images are used for training and 1176 images are used for testing. For the training of SharpNet, 4754 slices are used. Among them, 4742 images are used for training and 12 images are used for testing. Each slice has a size of 512×512 pixels. In the process of fan beam CT reconstruction, projection data is generated by applying Siddon's ray driven algorithm [38], [39]. The distances from the rotation center to both the source and the detector are fixed at 400mm. The detector size is set to 413 mm, with a detector element pitch of 0.8066 mm. The number of projection views is 512, and the number of detector elements is 512. The image pixel size is set to 0.8066 mm, with a matrix size of 512 × 512. Additionally, the total number of projection views is evenly distributed.

### B. Model Training and Parameter Selection

In our experiment, DMPD is trained by the AdamW algorithm with a learning rate of $2.0 \times 10^{-6}$. Samples are generated using DDIM sampling, where the shape of the sampled data is 2, 3, 128, 128, the eta value is set to 1.0, and the number of sampling steps is 250. The denoising network SharpNet is trained by the Adam algorithm, with a learning rate of $1.0 \times 10^{-3}$. Moreover, during the training process of the SharpNet, noise with a standard deviation of 0.1 is introduced to enable the model to learn how to handle noise. This method is implemented in Python using the operator discretization library (ODL) [40] and PyTorch on a personal workstation equipped with a GPU card NVIDIA 3090 24GB. In the generation stage, the first stage model conducts 72 steps of DDIM sampling. After the sampling is completed, the training weights are restored to complete the process of generating data in the projection domain. The second stage model denoises the image domain data after the FBP to complete the refinement process.

### C. Quantitative Indices

In order to evaluate the quality of the generated data, quantitative evaluations are conducted using the Frechet inception distance (FID), Inception Score (IS), and kernel inception distance (KID).

FID is an indicator used to evaluate the performance of generative models. It measures the quality of generated data by comparing the distribution differences between the generated data and the real data in the feature space. The following is the calculation formula:

$$\mathrm{FID} = \left\|\mu_r - \mu_g\right\|_2^2 + \mathrm{Tr}(\Sigma_r + \Sigma_g - 2(\Sigma_r \Sigma_g)^{1/2}) \tag{21}$$

where $\mu_r$ and $\mu_g$ are the mean vectors of real and generated data in the feature space, representing the data centers. $\Sigma_r$ and $\Sigma_g$ are the corresponding covariance matrices, describing variances and correlations of feature dimensions. $\|\cdot\|_2$ calculates the Euclidean distance between mean vectors to measure the difference in data centers. A lower FID value means the distribution of generated data is closer to that of real data, indicating higher generation quality of the model.

IS uses a pre-trained Inception network to calculate the class probability distribution of generated images. It computes two entropies and IS is the exponential difference of them as:

$$IS = \exp(H(p(y)) - \mathbb{E}_{x\sim p_g}\left[H(p(y\,|\,x))\right]) \tag{22}$$

where $H(p(\mathrm{y}))$ represents the overall distribution entropy of the generated images across all classes; $H(p(y\,|\,x))$ represents the conditional distribution entropy of the class to which the generated image x belongs; $p_g$ is the distribution of the generative model. The higher the value of the IS, the better the quality of the generated images.

KID uses kernel method to measure the distance between generated and real image distributions, computing kernel matrices and their traces for the value. KID is defined as:

$$\mathrm{KID} = (1/m)\sum_{i=1}^{m} k(x_i^g, x_i^g) + (1/n)\sum_{i=1}^{n} k(x_i^r, x_i^r) - (2/mn)\sum_{i=1}^{m}\sum_{j=1}^{n} k(x_i^g, x_j^r) \tag{23}$$

where $x_i^g$ and $x_i^r$ are the features of the generated images and the real images respectively, and m and n are the numbers of the generated images and the real images respectively. The lower the value of KID is, the closer the distribution of the generated images is to that of the real images, and the better the performance of the generative model is.

## D. *Image Domain VS. Projection Domain*

Image domain generation methods such as style-GAN [41] and DMT [42] synthesize CT images directly in pixel space, whereas the proposed PRO framework operates in the CT projection domain. As illustrated in Fig. 5, projection domain generation enables superior preservation of structural details and frequency information. For instance, in abdominal CT synthesis as shown in Fig. 5 (a), PRO captures not only external organ contours but also intricate internal structures. In contrast, image domain approaches often fail to maintain such anatomical fidelity, particularly in high frequency regions as shown in Fig. 5 (e).

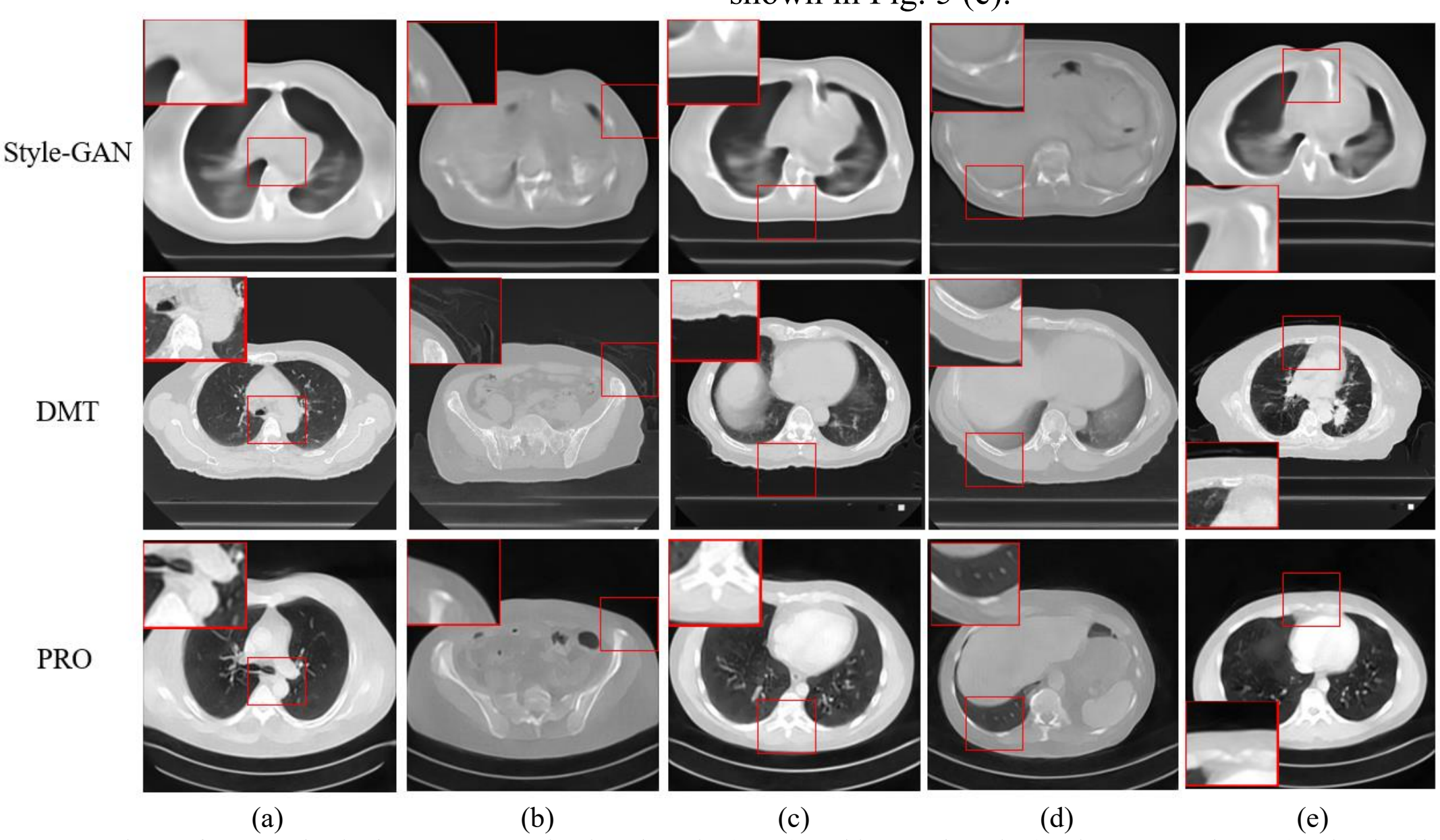

**Fig. 5.** Performance comparison of CT synthesis data among projection domain (PRO) and image domain (Style-GAN and DMT). The detailed magnified images are described in the upper left of each row. The lung region is presented with a window width of 1500 HU and a window level of -600 HU, while the abdominal region is displayed with a window width of 400 HU and a window level of 40 HU.

TABLE I
COMPARISON OF THE INDICATORS AMONG PRO, STYLE-GAN AND DMT FOR GENERATING 1000 CT IMAGES.

| Method | FID↓ | IS↑ | KID↓ |
|---|---|---|---|
| DMT | 0.9236 | 2.6812±0.1554 | 0.3039±0.0037 |
| Style-GAN | **0.4018** | 2.5072±0.1527 | 0.3497±0.0034 |
| PRO | 0.4241 | **3.0692±0.1492** | **0.3038±0.0028** |

Quantitative evaluations using FID, IS, and KID further demonstrate the advantages of PRO. As summarized in Table I, PRO achieves the second lowest FID score, indicating higher visual fidelity. Its IS score reaches 3.0692, outperforming DMT 2.6812 and style-GAN 2.5072, while maintaining a lower standard deviation 0.1492, reflecting improved generative stability. Here, the symbol "↓" indicates that a lower value of the indicator is better, while "↑" means a higher value is better. In terms of distribution consistency, PRO achieves the lowest KID mean value 0.0028, compared to DMT 0.0037 and style-GAN 0.0034. These results collectively validate the effectiveness of projection domain generation in producing high quality, structurally consistent CT images for medical applications.

## E. *Different Generative Methods in Projection Domain*

PRO is compared with the current mainstream generative methods, including diffusion-based [43] method and GAN-based [44] method.

To systematically evaluate the generation capabilities of different models in the CT projection domain, we conducted a comparative analysis for PRO, a GAN-based method, and a diffusion-based method. In terms of generation in the projection domain, the richer the details presented in the chord diagram, the stronger the corresponding method's generation ability. As shown in Fig. 6, PRO method outperforms both the diffusion-based and GAN-based methods. The GAN network is highly prone to overfitting during the training process, resulting in poor diversity and detail performance of the generated projection domain data.

On the other hand, when the diffusion-based method generates projection domain data, its insufficient ability to learn from noise leads to the failure of completely eliminating the noise. In contrast, PRO achieves the best results in both the diversity of the generated projection domain data and the consistency of details, which provides extremely powerful support for the subsequent two stage image generative model. To further validate the effectiveness of the multi-stage model in CT image generation, PRO is compared against diffusion-based and GAN-based methods. As illustrated in Fig. 7, PRO outperforms the others significantly. The GAN-based approach, hampered by its limited projection domain generation diversity and highly unstable training process, produces CT images lacking substantial details. Meanwhile, the diffusion-based method struggles to effectively remove noise due to the diffusion model's suboptimal handling of projection domain interference, resulting in generated images with prominent artifacts and severe detail loss. In contrast, PRO excels in both projection domain generation and image refinement, achieving high fidelity CT image generation.

To quantify the performance of different methods in generating image domain data from CT projection data, we compare PRO with GAN-based and diffusion-based methods across multiple metrics. The results, summarized in Table II, evaluate the quality, diversity, and stability of generated images using 100 corresponding image pairs. Regarding the FID, a lower value implies that the distribution of the generated data is closer to that of the real data, which means higher quality of the generated data. Compared with the GAN-based and diffusion-based methods, PRO achieves the best performance in terms of the quality of the generated data. For the IS, a higher mean value indicates better quality and greater diversity of the generated images. As shown in Fig. 6, due to the poor performance of GAN and diffusion-based in the projection domain, the generated images lack diversity, resulting in a lower mean IS value. In contrast, PRO outperforms both the GAN and diffusion-based methods in terms of both the quality and diversity of the generated images. A lower mean value of the KID indicates that the feature distribution of the generated image domain more closely resembles that of real images. A lower standard deviation means that the degree of closeness between the distribution of the generated images and that of the real images is relatively stable when the generative model generates images. PRO can not only generate higher quality images but also significantly enhance the diversity of the generated images.

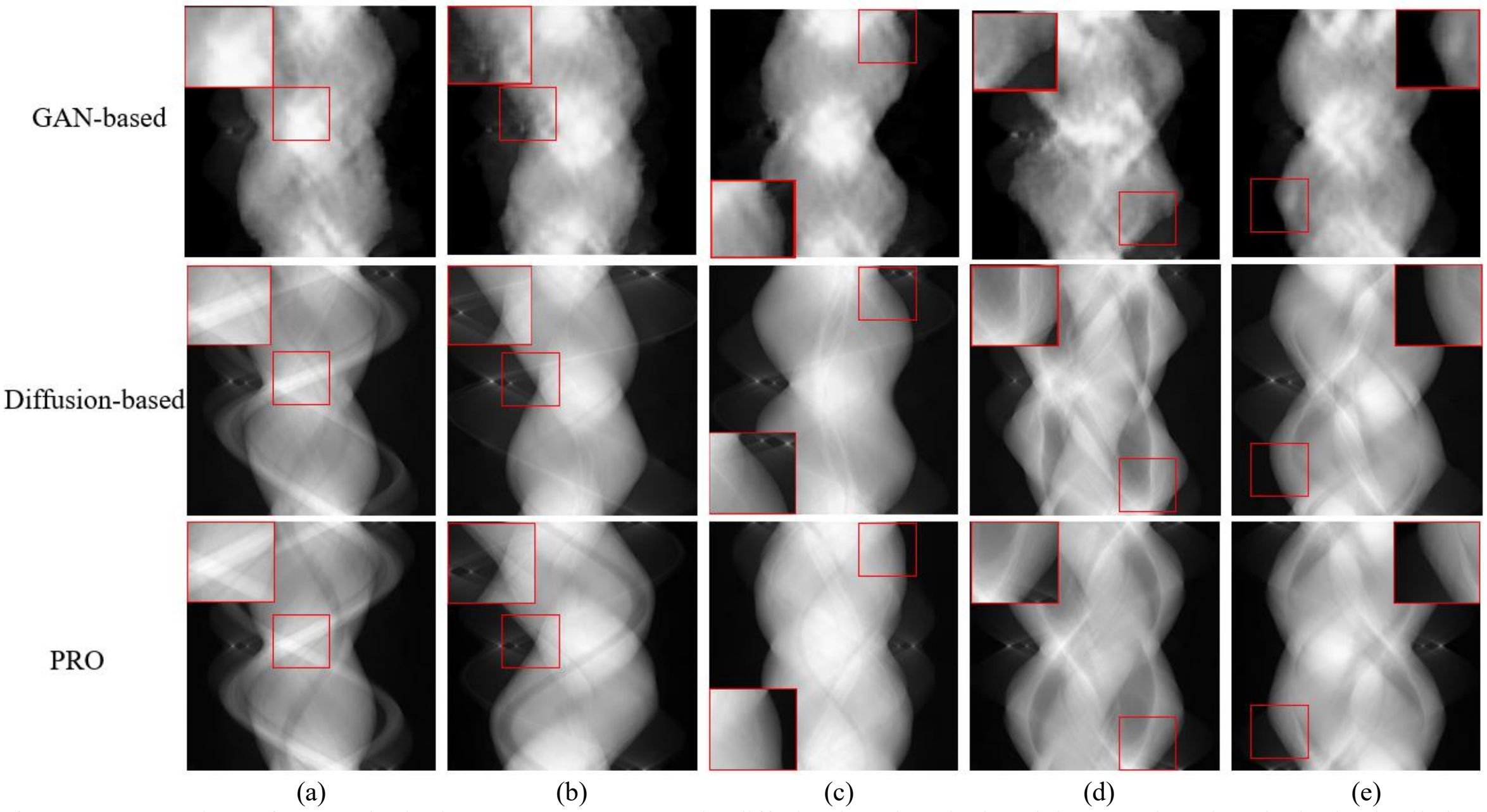

**Fig. 6.** Performance comparison of CT synthesis sinograms among PRO, the diffusion-based method, and the GAN-based method. The detailed magnified images are described in the upper left of each row.

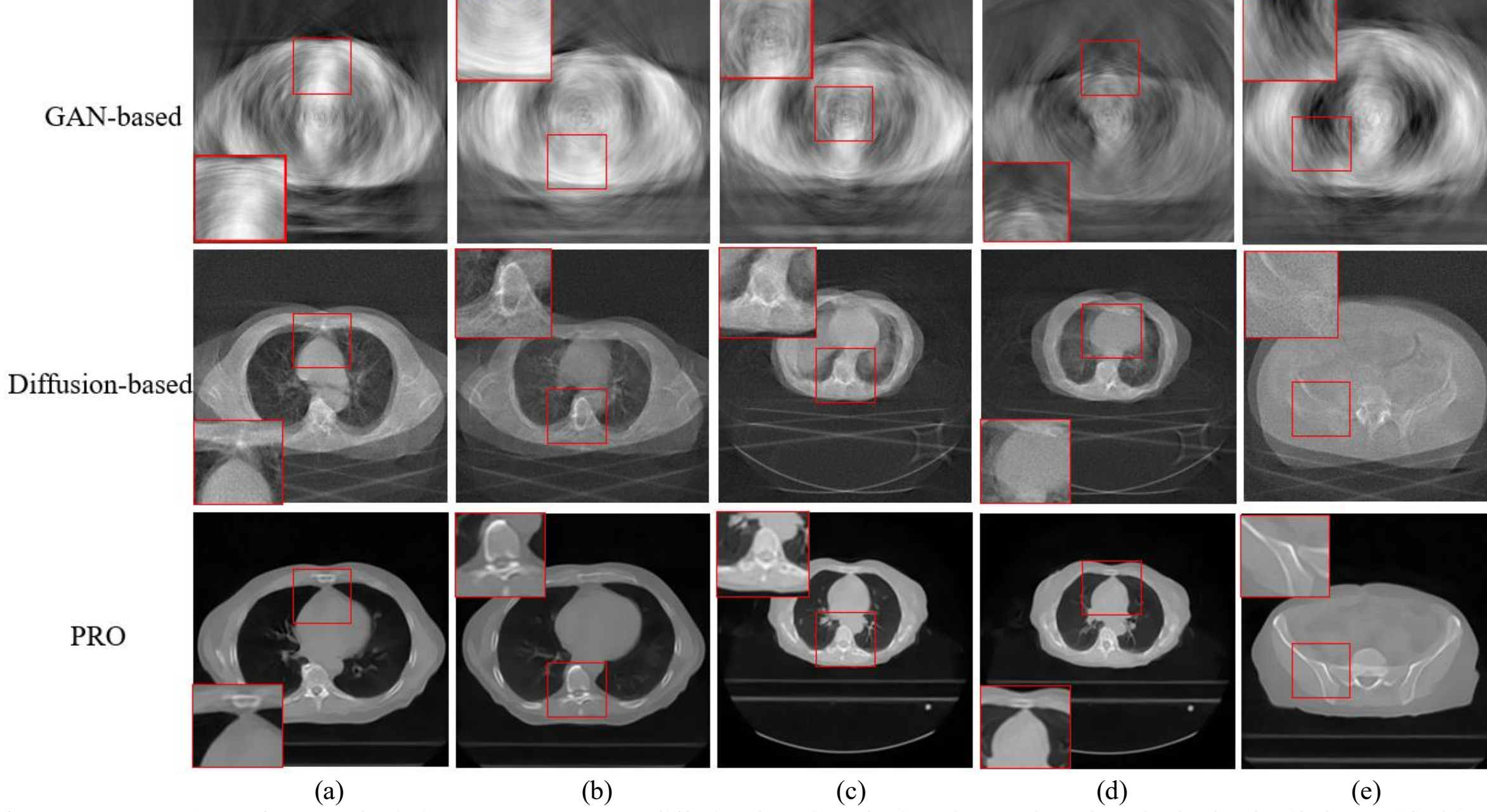

**Fig. 7.** Performance comparison of CT synthesis images among PRO, diffusion-based method, and GAN-based method. The detailed magnified images are described in the upper left of each row.

To comprehensively assess the scalability and consistency of generative models in large scale CT image synthesis, we conducted experiments generating 1,000 CT images using PRO, GAN-based, and diffusion-based approaches in Table II. In terms of the FID metric, PRO achieves a value of 0.4241, which is significantly lower than the 2.2506 of the GAN-based method and the 2.2155 of the diffusion-based method. This indicates that the CT images generated by PRO are closer to the real data in terms of distribution characteristics, and the generation quality is notably superior to the comparison methods. Regarding the IS metric, the IS mean of PRO is 3.0692, which is much higher than the 1.0050 of the GAN-based method and the 1.0119 of the diffusion-based method. Moreover, the IS std is only 0.1492, suggesting that PRO can not only generate high quality and diverse images but also ensure the stability of the generation results. In contrast, the low IS mean values of the GAN-based and diffusion-based methods indicate obvious deficiencies in the quality and diversity of the generated images. The KID mean of PRO is 0.3038, lower than the 1.0299 of the GAN-based method and the 1.0267 of the diffusion-based method. Combined with the KID std of only 0.0028, it further validates the advantages of PRO in feature distribution matching and generation stability. Although the GAN-based and diffusion-based methods have relatively low KID mean values, their distribution stability is still inferior to PRO when considering the standard deviation.

In Table II, whether generating 100 or 1000 CT images, PRO significantly outperforms the GAN-based and diffusion-based methods in the FID, IS, and KID metrics. It can not only generate higher quality and more diverse CT images but also ensure the stability and reliability of the generation process, fully verifying the effectiveness and superiority of the proposed method in the medical image generation task.

TABLE II

COMPARISON OF THE INDICATORS AMONG PRO, THE DIFFUSION-BASED METHOD AND THE GAN-BASED METHOD FOR GENERATING 100 CT IMAGES.

| | Generating 100 CT Images | | | Generating 1000 CT Images | | |
|---|---|---|---|---|---|---|
| | FID | IS | KID | FID | IS | KID |
| GAN-based | 2.2515 | 1.0039±**0.0034** | 1.0455±0.0543 | 2.2506 | 1.0050±0.0016 | 1.0299±0.0544 |
| Diffusion-based | 2.2154 | 1.1471±0.0227 | 0.5647±0.0145 | 2.2155 | 1.0119±**0.0013** | 1.0267±0.0165 |
| PRO | **0.4425** | **2.4579**±0.3149 | **0.3030±0.0094** | **0.4241** | **3.0692**±0.1492 | **0.3038±0.0028** |

## F. *Ablation Study*

Diffusion model has demonstrated impressive capabilities in image generation. However, generating data directly in the projection domain presents unique challenges. Small variations in the sinogram can result in significant distortions in the reconstructed CT images, making it difficult for diffusion model to completely remove noise during projection domain generation. To address this limitation, we propose a two stage framework. The first stage, DMPD is PRO w/o SharpNet, performs the initial sinogram synthesis. The second stage, SharpNet, operates in the image domain to refine reconstructed images, enhancing structural clarity and recovering fine details.

Comparative evaluations are conducted among DMPD and PRO as shown in Fig. 8. Visual results indicate that DMPD effectively reduces large scale artifacts and preserves global anatomical structures. SharpNet further improves the local image quality by suppressing residual noise and enhancing textures, resulting in clearer and more detailed outputs.

Quantitative results on generating 1000 CT images demonstrate that the proposed two stage approach offers superior performance. The PRO framework achieves an FID of 0.4241, indicating improved alignment with real data compared to the 0.4921 achieved by DMPD alone. The IS for PRO reaches 3.0692, surpassing the 2.9747 obtained by DMPD, and shows a lower standard deviation, reflecting greater consistency. While the mean KID of PRO is marginally higher than that of DMPD, its standard deviation is substantially smaller, indicating improved reliability across samples. These findings confirm that the integration of projection domain generation with image domain refinement leads to enhanced image fidelity, stability, and overall synthesis quality.

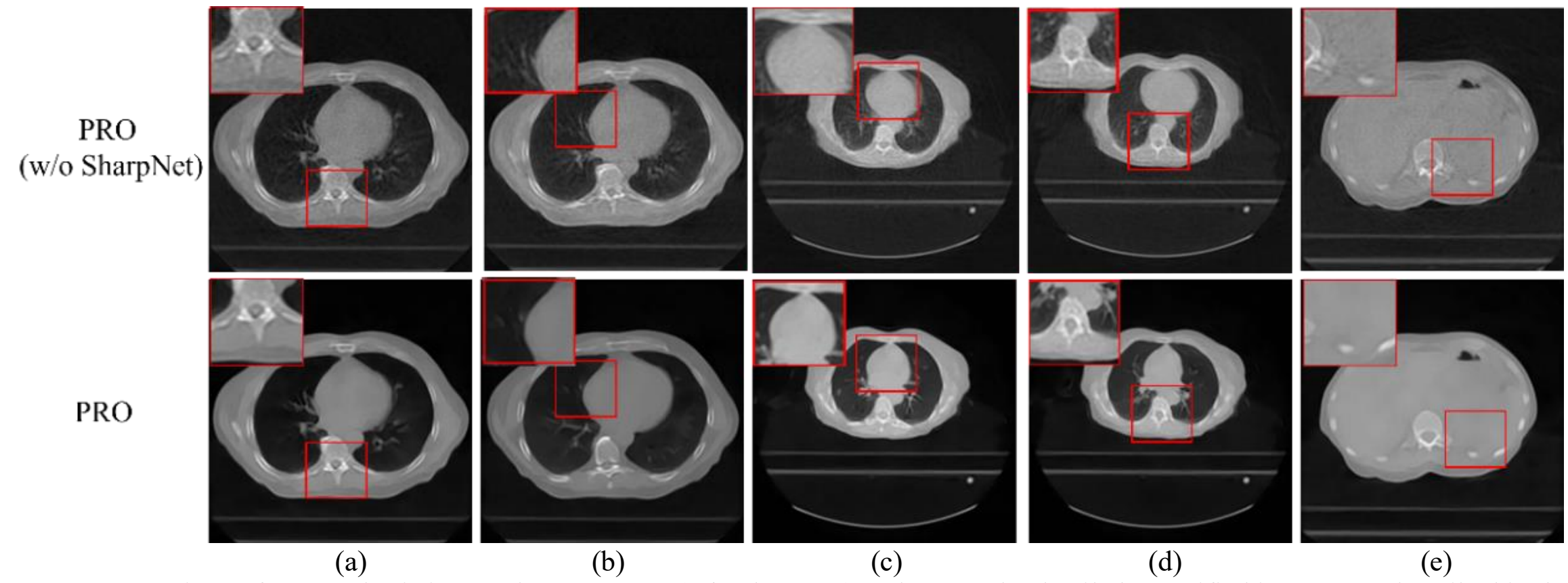

**Fig. 8.** Performance comparison of CT synthesis images between PRO(w/o SharpNet) and PRO. The detailed magnified images are described in the upper left of each row.

Two latent space configurations of PRO are compared using VQ encoders with downsampling factors of 4× and 8×. The 4× configuration, with deeper hierarchical compression, captures more global features and long range dependencies, which are essential for maintaining the structural integrity of projection domain CT synthesis. In contrast, the 8× configuration shows limitations in global representation due to lower spatial granularity.

As reported in Table IV, the 4× configuration achieves an FID of 0.4425, lower than the 0.6520 of 8×. The IS standard deviation for 4× is 0.3149, indicating more stable generation, while 8× reaches 0.3848. Although the IS mean of 8× is slightly higher, 4× demonstrates better overall consistency. For KID, 4× attains a lower mean of 0.3030 and a reduced standard deviation of 0.0094, confirming stronger feature alignment and reliability.

These results suggest that the 4× latent space offers superior performance across fidelity, stability, and feature similarity. It provides a high quality representation for projection domain generation and enhances input quality for the image domain SharpNet stage, contributing to improved CT synthesis throughout the full pipeline.

TABLE III
COMPARISON OF THE INDICATORS AMONG DMPD AND PRO FOR GENERATING 1000 CT IMAGES.

| | PRO (w/o SharpNet) | PRO |
|---|---|---|
| FID | 0.4921 | **0.4241** |
| IS | 2.9747±0.2036 | **3.0692±0.1492** |
| KID | **0.2933**±0.0108 | 0.3038±**0.0028** |

TABLE IV
COMPARISON FOR 100 IMAGES GENERATED WITH DIFFERENT DOWNSAMPLING FACTORS.

| | 8× Downsampling | 4× Downsapling (PRO) |
|---|---|---|
| FID | 0.6520 | **0.4425** |
| IS | **2.7046**±0.3848 | 2.4579±**0.3149** |
| KID | 0.3565±0.0105 | **0.3030±0.0094** |

### G. *Synthetic Datasets for Different Tasks*

PRO is employed to synthetic 2000 CT datasets. These generated datasets are then used to replace the AAPM datasets, achieving comparable performance in low-dose and sparse-view tasks. In the sparse-view CT reconstruction task, GMSD (AAPM) is trained using 2,000 datasets randomly selected from the original 4,742 AAPM dataset. Conversely, we employ 2,000 synthetic datasets generated by PRO for GMSD (PRO) training. Comparative results are presented in Table V show that PRO with AAPM data at views 60, 90, 120 and 180. PRO achieves 37.32 in PSNR, 0.9114 in SSIM, and 0.0002 in MSE, slightly surpassing AAPM data which yields 37.15, 0.9111, and 0.0002 respectively in view 90. PRO provides an increase of 0.17 in PSNR and a marginal gain of 0.0003 in SSIM while maintaining the same MSE. These findings demonstrate the authenticity of synthetic data by PRO.

TABLE V
APPLICATION PSNR/SSIM/MSE OF PRO GENERATED DATA IN SPARSE-VIEW CT RECONSTRUCTION METHOD GMSD.

| Views | GMSD (AAPM) | GMSD (PRO) |
|---|---|---|
| 60 | **34.905**/0.8999/0.0003 | 34.417/**0.9044/0.0003** |
| 90 | 37.150/0.9111/0.0002 | **37.321/0.9114/0.0002** |
| 120 | **39.897/0.9586**/0.0001 | 38.792/0.9481/**0.0001** |
| 180 | **42.091/0.9706**/0.0001 | 40.586/0.9625/**0.0001** |

For low-dose CT reconstruction task, OSDM (AAPM) is trained by AAPM dataset. In contrast, OSDM (PRO) is trained using 2,000 synthetic datasets generated by PRO. A comparison of the results is presented in Table VI at a noise level of 1e5, 5e4 and 1e4. PRO achieves 41.74 in PSNR, 0.9670 in SSIM, and 0.0001 in MSE at a noise level of 5e4, whereas AAPM data produces 41.20, 0.9857, and 0.0001. The PSNR improved by 0.54, demonstrating that image quality can be maintained even when radiation dose is reduced.

TABLE VI
APPLICATION PSNR/SSIM/MSE OF PRO GENERATED DATA IN LOW-DOSE CT RECONSTRUCTION METHOD OSDM.

| Noise Level | OSDM (AAPM) | OSDM (PRO) |
|---|---|---|
| 1e5 | **42.62/0.9899**/0.0001 | 42.49/0.9777/**0.0001** |
| 5e4 | 41.20/**0.9857**/0.0001 | **41.74**/0.9670/**0.0001** |
| 1e4 | **37.43/0.9683/0.0002** | 34.56/0.9242/0.0004 |

Together, these results highlight the effectiveness of PRO in leveraging projection domain priors to generate synthetic data that supports high quality reconstruction. By generalizing across tasks such as sparse-view and low-dose imaging, PRO exhibits the key characteristics of a foundation model for CT, including strong adaptability, task scalability, and the ability to deliver consistent performance across varying clinical conditions.

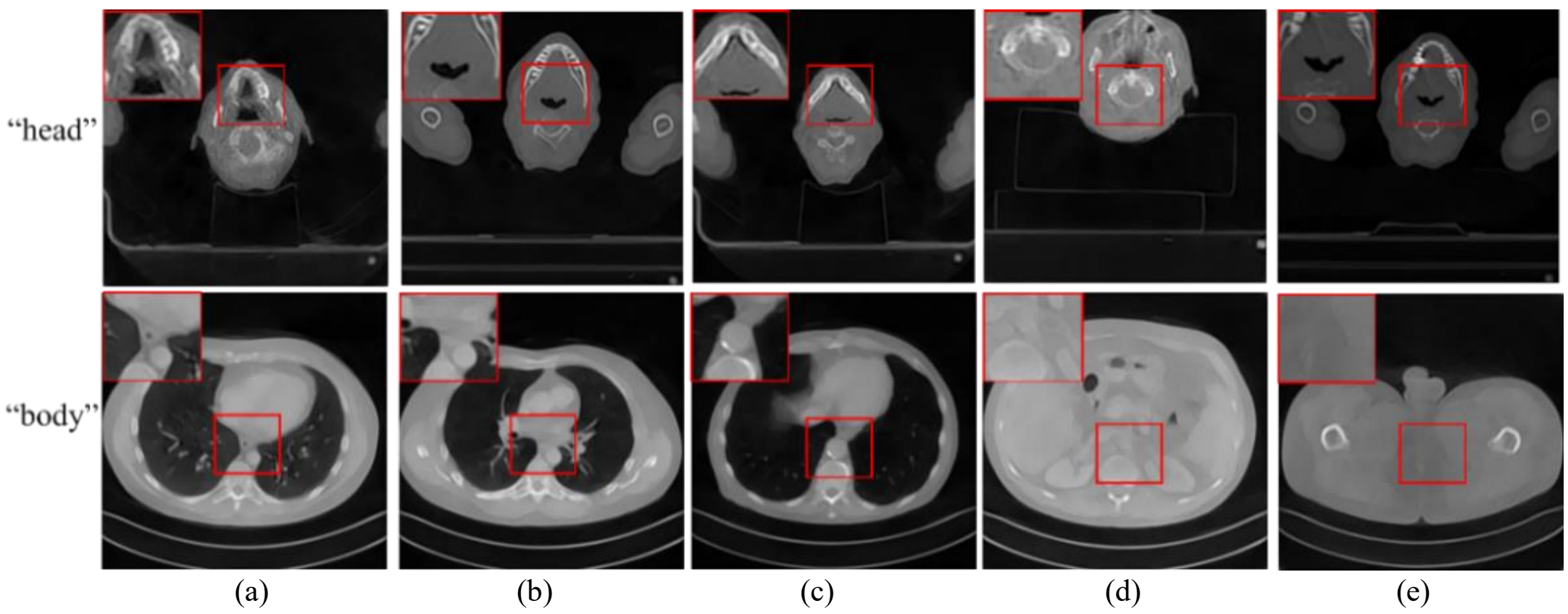

**Fig. 9.** CT synthesis images with task specified texts for “head” and “body”. The detailed magnified images are described in the upper left of each row.

## V. DISCUSSION

The proposed PRO foundation model enables text conditioned sinogram generation tailored to specific anatomical regions and clinical tasks. During training, anatomical prompts such as “head” and “body” guide the model to encode distinct structural priors within separate latent subspaces. For instance, the model is trained with 500 head region CT images and validated on 100 held out images from the AAPM dataset under the “head” prompt. As shown in Fig. 9, the synthesized head CT outputs exhibit clear delineation of cranial structures, validating the model’s capacity to learn region specific anatomical representations. Similarly, for the “body” prompt, PRO is trained on 4,000 abdominal images and evaluated on 742 AAPM samples. The generated sinograms yield reconstructions with comparable fidelity and internal organ consistency, demonstrating the model’s ability to generalize across diverse anatomical zones while retaining structural plausibility.

While PRO has demonstrated strong capabilities in synthesizing anatomically plausible CT sinograms for head and body regions via text prompts, future research could explore the integration of more diverse and fine grained task prompts. For instance, incorporating prompts such as “metal artifact”, “beam hardening”, or “respiratory motion” would enable the model to generate physically realistic sinograms for challenging reconstruction scenarios. This approach may also assist in benchmarking and training artifact correction or robust reconstruction networks under controlled conditions. In addition, the proposed text conditioned generation framework can be extended to support the synthesis of paired data, including projection label pairs or artifact clean references. This would significantly benefit supervised learning tasks, such as metal artifact reduction, segmentation, or quality assessment, especially in domains where annotated clinical data is scarce. Ultimately, such capability could promote a task driven data centric paradigm for CT imaging, where high quality synthetic datasets can be generated on demand for targeted clinical applications.

## VI. Conclusions

In this study, we proposed PRO, a projection domain foundation model for CT image synthesis, which addressed data scarcity and complex imaging physics by directly modeling projection data and integrating task specific prompts for controllable synthesis. DMPD generated in the projection domain to obtain more prior information, while SharpNet refined the image domain after FBP to capture more image details. Experimental results demonstrated that CT data generated by PRO not only achieved more realistic effects but also enhanced the performance of downstream tasks such as low-dose and sparse-view CT reconstruction, verifying the effectiveness of projection domain generated data. This work highlighted projection domain synthesis as a foundation model for CT data generation and augmentation. Future directions include expanding the framework in several ways. First, it can be extended to support more diverse generative architectures. Second, multi-modal inputs such as MRI and PET can be integrated into the model. Third, the prompt vocabulary can be enriched with clinically meaningful terms. These improvements will further support a task driven and data centric paradigm for medical imaging.

## Acknowledgments

All authors declare that they have no known conflicts of interest in terms of competing financial interests or personal relationships that could have an influence or are relevant to the work reported in this paper.

## Appendix

### *A. CT Projection Data Generated by PRO*

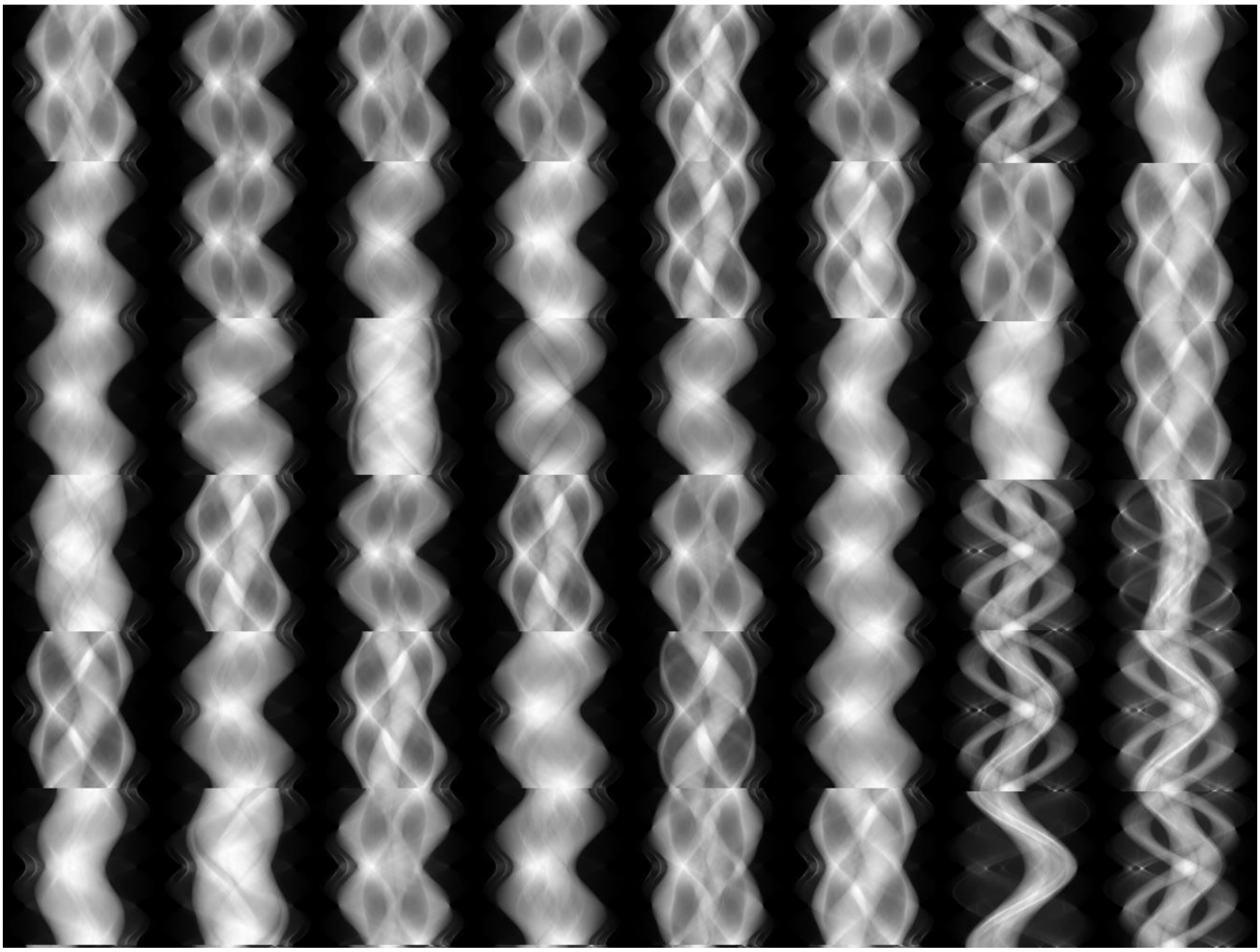

**Fig. 10.** Random generated samples of PRO on CT projection data with “body” and “brain” prompts. Sampled with 72 DDIM steps.

### B. CT Image Data Generated by PRO

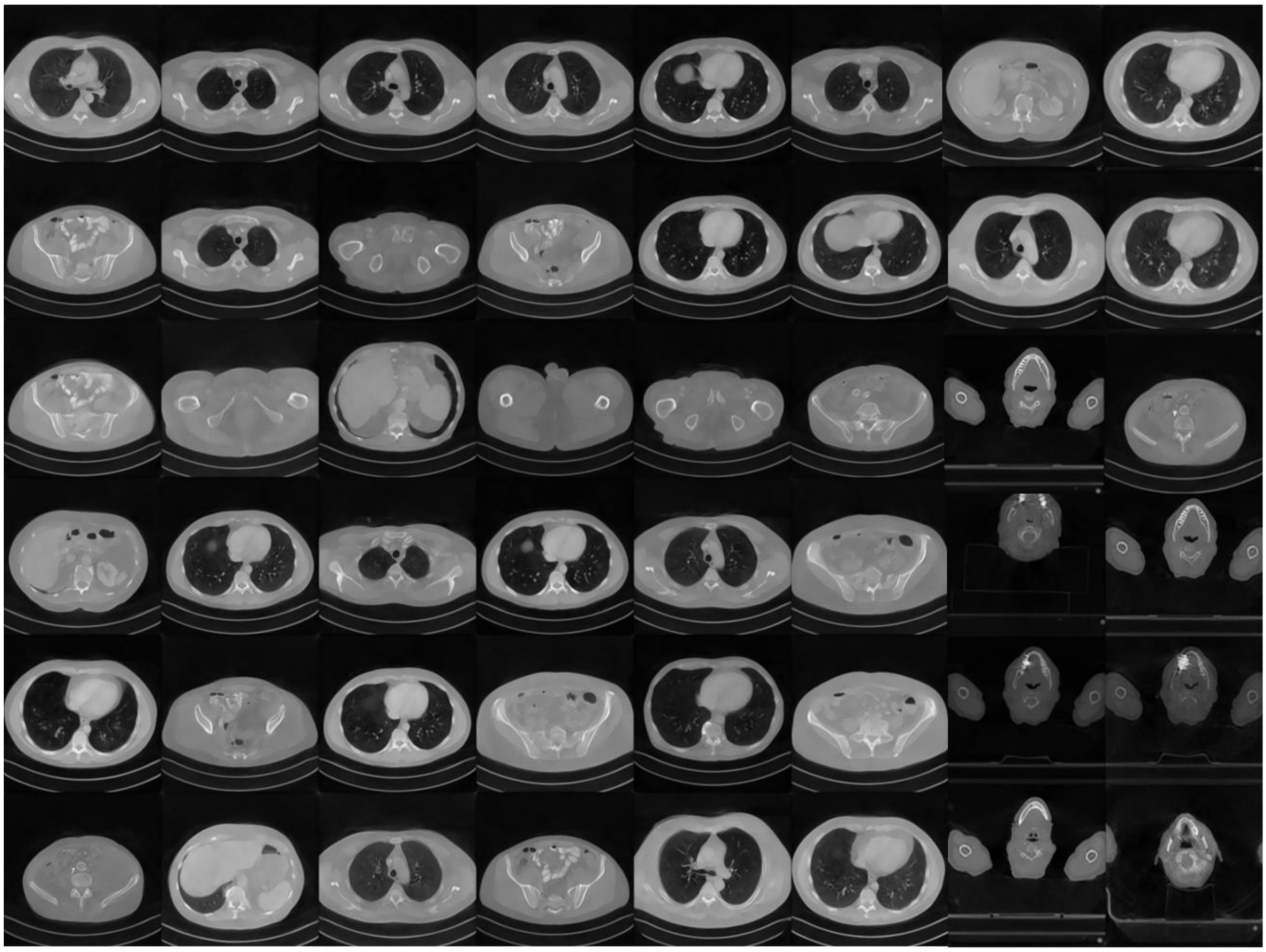

**Fig. 11.** Random generated samples of PRO on CT image data with "body" and "brain" prompts. Sampled with 72 DDIM steps.

## REFERENCES

[1] T. Zhou, S. Ruan, and S. Canu, *et al*, "A review: deep learning for medical image segmentation using multi-modality fusion," *Array*, 3: 100004, 2019.

[2] M. Yaqub, F. Jinchao, and K. Arshid, *et al*, "Deep learning-based image reconstruction for different medical imaging modalities," *Comput. Math. Methods Med.*, 2022(1): 8750648, 2022.

[3] M. T. Tran, S. H. Kim, and H. J. Yang, *et al*, "Medical image inpainting with deep neural network," *Proc. Smart Media Spring Conf.*, pp. 22-23, 2020.

[4] J. Y. Zhu, T. Park, and P. Isola, *et al*, "Unpaired image-to-image translation using cycle-consistent adversarial networks," *ICCV*, pp. 2223-2232, 2017.

[5] J. Jordon, L. Szpruch, and F. Houssiau, *et al*, "Synthetic data-what, why and how?," arXiv preprint arXiv:2205.03257, 2022.

[6] J. Jordon, A. Wilson, and M. Schaar, *et al*, "Synthetic data: Opening the data floodgates to enable faster, more directed development of machine learning methods," arXiv preprint arXiv:2012.04580, 2020.

[7] T. Wang, Y. Lei, and Y. Fu, *et al*, "A review on medical imaging synthesis using deep learning and its clinical applications," *JACMP.*, vol. 22, no. 1, pp. 11–36, 2021.

[8] A. Creswell, T. White, and A. A. Bharath, *et al*, "Generative adversarial networks: An Overview," *IEEE Signal Process. Mag.,* vol. 35, no. 1, pp. 53-65, Jan. 2018.

[9] L. Wang, W. Chen, and F. R. Yu, *et al*, "A state-of-the-art review on image synthesis with generative adversarial networks," *IEEE Access*, vol. 8, pp. 63514-63537, 2020.

[10] G. Kwon, C. Han, and D. Kim, *et al*, "Generation of 3D brain MRI using auto-encoding generative adversarial networks," *MICCAI. Cham: Springer International Publishing*, pp. 118-126, 2019.

[11] L. Sun, J. Chen, and K. Batmanghelich, *et al*, "Hierarchical amortized GAN for 3D high resolution medical image synthesis," *IEEE JBHI.*, vol. 26, no. 8, pp. 3966-3975, Aug. 2022, doi: 10.1109/JBHI.2022.3172976.

[12] N. Kodali, J. Abernethy, and J. Hays, *et al*, "On convergence and stability of gans," arXiv preprint arXiv:1705.07215, 2017.

[13] R. Rombach, A. Blattmann, and D. Lorenz, *et al*, "High-resolution image synthesis with latent diffusion models," *CVPR.*, pp. 10684-10695, 2022.

[14] W. H. L. Pinaya, P. D. Tudosiu, and J. Dafflon, *et al*, "Brain imaging generation with latent diffusion models," *MICCAI*. Cham: Springer Nature Switzerland, 2022, pp. 117-126.

[15] P. Guo, C. Zhao, and D. Yang, *et al*, "MAISI: Medical AI for synthetic Imaging," *WACV.*, Tucson, AZ, USA, 2025, pp. 4430-4441.

[16] I. E. Hamamci, A. Tezcan, and A. G. Simsek, *et al*, "Generatect: Text-guided 3d chest ct generation," *CoRR*, 2023.

[17] B. Guan, C. Yang, and L. Zhang, *et al*, "Generative modeling in sinogram domain for sparse-view CT reconstruction," *IEEE TRPMS.*, vol. 8, no. 2, pp. 195-207, 2023.

[18] B. Huang, S. Lu, and L. Zhang, *et al*, "One-sample diffusion modeling in projection domain for low-dose CT imaging," *IEEE TRPMS.*, vol. 8, no.8, pp. 902-915, 2024.

[19] W. Zhang, B. Huang, and S. Chen, *et al*, "Low-rank angular prior guided multi-diffusion model for few-shot low-dose CT reconstruction," *IEEE Trans. Comput. Imaging.*, vol. 10, pp. 1763-1774, 2024.

[20] K. Xu, S. Lu, and B. Huang, *et al*, "Stage-by-stage wavelet optimization refinement diffusion model for sparse-view CT reconstruction," *IEEE Trans. Med. Imag.*, vol. 43, no. 10, pp. 3412-3424, 2024.

[21] A. Radford, L. Metz, and S. Chintala, *et al*, "Unsupervised representation learning with deep convolutional generative adversarial networks," *ICLR*, 2016.

[22] A. Oord, S. Dieleman, and O. Vinyals, *et al*, "Wavenet: A generative model for raw audio," arXiv preprint arXiv:1609.03499, 2016.

[23] S. R. Bowman, L. Vilnis, and S. Bengio, *et al*, "Generating sentences from a continuous space," *CONLL*, pp. 1021, 2016.

[24] D. P. Kingma and M. Welling, "Auto-encoding variational Bayes," *ICLR*, 2014.

[25] K. C. Tezcan, C. F. Baumgartner, and E. Konukoglu, *et al*, "MR image

reconstruction using deep density priors," *IEEE Trans. Med. Imag.*, vol. 38, no. 7, pp. 1633-1642, 2018.

[26] T. Temizel, Tugba, and C. Matthew, "A comparative study of autoregressive neural network hybrids," *Neural Networks*, vol. 18, no. 5-6, pp. 781-789, 2005.

[27] M. Asim, A. Ahmed, and P. Hand, "Invertible generative models for inverse problems: mitigating representation error and dataset bias," *ICML*, pp. 399-409, 2020.

[28] Y. Song, S. Ermon, "Improved techniques for training score-based generative models," *Advances in Neural Information Processing Systems*, vol. 33, pp. 12438-12448, 2020.

[29] R. Salakhutdinov and G.E. Hinton, "Deep Boltzmann machines," *Proc. Int. Conf. Artif. Intell. Statist.*, pp. 448-455, 2009.

[30] A. Graves, "Generating sequences with recurrent neural networks," arXiv preprint arXiv:1308.0850, 2013.

[31] E. L. Denton, S. Chintala, and R. Fergus, "Deep generative image models using a Laplacian pyramid of adversarial networks," *Adv. Neural Inf. Process. Syst.*, pp. 1486-1494, 2015.

[32] P. Vincent, "A connection between score matching and denoising autoencoders," *Neural computation*, vol. 23, no. 7, pp. 1661-1674, 2011.

[33] G. Parisi, "Correlation functions and computer simulations," *Nuclear Physics*, vol. 180, no. 3 pp. 378-384, 1981.

[34] Y. Song, S. Ermon, "Generative modeling by estimating gradients of the data distribution," *Advances in Neural Information Processing Systems*, vol. 32, 2019.

[35] S. Banerjee, E. Jiaze, and B. Ren, *et al*, "Inpainting the sinogram from computed tomography using latent diffusion model and physics," *ICLR*, 2025.

[36] Z. Liu, R. Kettimuthu, and I. Foster, "Masked sinogram model with transformer for ill-posed computed tomography reconstruction: a preliminary study," arXiv preprint arXiv:2209.01356, 2022.

[37] Low Dose CT Grand Challenge. Accessed: Apr. 6, 2017. [Online]. Available: http://www.aapm.org/GrandChallenge/LowDoseCT/.

[38] R. L. Siddon, "Fast calculation of the exact radiological path fora three-dimensional CT array," *Medical Physics*, vol. 12, no. 2, pp.252 - 255, 1985.

[39] F. Jacobs, E. Sundermann, and I. Lemahieu, *et al*, "A fast algorithm to calculate the exact radiologi-calpath through a pixel or voxel space," *Journal of Computing and Information Technology*, vol. 6, no. 1, pp. 89 - 94, 1998.

[40] J. Adler, H. Kohr, and O. Oktem, "Operator discretization library (ODL)," Software available from https://github.com/odlgroup/odl, vol. 5, 2017.

[41] Y. Song, L. Shen, and L. Xing, *et al*, "Solving inverse problems in medical imaging with score-based generative models," arXiv preprint arXiv:2111.08005, 2021.

[42] L. Fetty, M. Bylund, and P. Kuess, *et al*, "Latent space manipulation for high-resolution medical image synthesis via the StyleGAN," *Zeitschrift für Medizinische Physik*, vol. 30, no. 4 pp. 305-314, 2020.

[43] S. Pan, T. Wang, and R. L. J. Qiu, *et al*, "2D medical image synthesis using transformer-based denoising diffusion probabilistic model," *Physics in Medicine & Biology*, v0l. 68, no. 10 pp. 105004,2023.

[44] T. Karras, S. Laine, and M. Aittala, *et al*, "Analyzing and improving the image quality of stylegan" *CVPR.*, pp. 8110-8119, 2020.